\documentclass[11pt,reqno]{amsart}

\usepackage{amsfonts}
\usepackage{mathrsfs}
\allowdisplaybreaks[4]
\usepackage[dvips]{graphicx} 
\usepackage{epsfig}
\usepackage{amssymb}
\usepackage{epsfig}
\usepackage{amsmath}
\usepackage{cite}

\newcommand{\be}{\begin{equation}}
\newcommand{\ee}{\end{equation}}
\newcommand{\bea}{\begin{eqnarray}}
\newcommand{\eea}{\end{eqnarray}}
\newcommand{\ba}{\begin{array}}
\newcommand{\ea}{\end{array}}
\newcommand{\bean}{\begin{eqnarray*}}
\newcommand{\eean}{\end{eqnarray*}}

\begin{document}
\title[Integrability ]{two kinds of rogue waves of the general nonlinear Schr\"odinger equation with derivative }
\author{Shuwei Xu\dag,Jingsong He$^*$ \dag,Lihong Wang \dag
 }

 \thanks{$^*$ Corresponding author: hejingsong@nbu.edu.cn;jshe@ustc.edu.cn}
 \maketitle
\dedicatory {  \dag \ Department of Mathematics, Ningbo University,
Ningbo , Zhejiang 315211, P.\ R.\ China\\}

\begin{abstract}
In this letter,the designable integrability(DI) of the variable
coefficient derivative nonlinear Schr\"odinger equation (VCDNLSE) is
shown by construction of an explicit transformation which maps
VCDNLSE to the usual derivative nonlinear Schr\"odinger
equation(DNLSE). One novel feature of VCDNLSE with DI is that its
coefficients can be designed artificially and analytically by using
transformation. What is more, from the rogue wave and rational
traveling solution of the DNLSE, we get two kinds of rogue waves of
the VCDNLSE by this transformation. One kind of rogue wave has
vanishing boundary condition, and the other non-vanishing boundary
condition.  The DI of the VCDNLSE also provides a possible way to
control the profile of the rogue wave in physical experiments.
\end{abstract}

{\bf PACS(2010) numbers}: 02.30.Ik, 42.81.Dp, 52.35.Bj, 52.35.Sb, 94.05.Fg


\maketitle
\section{Introduction}
The derivative nonlinear Schr\"odinger equation (DNLSE) is
originated from two fields of applied physics. Firstly, the DNLSE
governs the evolution of small but finite amplitude Alfv\'en waves
propagating quasi-parallel to the magnetic field \cite{MDNLS} in
plasma physics. Recently, this equation has also been used to describe
rogue waves \cite{ruderman2,derman2,xuhe1,Moslem,Laveder} in
plasmas.Secondly, nonlinear optics, the sub-picosecond or
femtosecond pulses in single-mode optical fiber\cite{WDNLS,anderson}
is modeled by the DNLSE. There exist three kinds of DNLSE, i.e,. the
DNLSIE or DNLSE \cite{KN}
\begin{equation}\label{dnlsI}
iq_{t}-q_{xx}+i(q^2q^\ast)_{x}=0,
\end{equation}
the DNLSIIE \cite{CL}
\begin{equation}\label{CL}
iq_{t}+q_{xx}+iqq^\ast q_{x}=0,
\end{equation}
and the DNLSIIIE\cite{GI}
\begin{equation}\label{GI}
iq_{t}+q_{xx}-iq^2{q_{x}}^{\ast}+\frac{1}{2}q^3{q^\ast}^2=0,
\end{equation}
and a chain of gauge transformations between them:
DNLSIIE $\stackrel{a)}{\Longrightarrow}$
 DNLSIE $\stackrel{b)}{\Longrightarrow}$ DNLSIIIE.
Here a) denotes eq.(2.12) in ref.\cite{wadati1}, and b) denotes: eq.(4)$\rightarrow$ eq.(3)
$\rightarrow$eq.(6) with $\gamma=0$ in ref.\cite{kakei1}. But these transformations can't preserve
the reduction conditions in spectral problem of the Kaup and Newell (KN) \cite{KN} system and involve complicated integrations.
So each of them deserves to be investigated separately. The goal of the present work is to extend consideration to the DNLSE.

  Most known works on the DNLSE equations focus  exclusively on equations with constant
  coefficients\cite{xuhe1,KN,kenji1,steduel,kawata1,huang2}. For wave
propagation through an inhomogeneous medium, the variable
coefficient derivative nonlinear Schr\"odinger equation (VCDNLSE)
\cite{Buti2,Chow,tian1} will arise. In fact the nonlinear
Schr\"odinger equation  with variable
coefficients\cite{liu,Belmonte2,Serkin2,Kundu,yan,wu,he} have
already been studied extensively in the literature. A simple and
relevant example will be the properties of short optical pulses
along a fiber with spatially increasing/decreasing dispersion. A
variety of ingenious similarity transformations\cite{he} has been
introduced. The solitary pulse can undergo broadening or
compression, with or without the presence of damping/gain. One
objective here is to determine if these phenomena will continue to
exist for the DNLSE.The analytical technique employed here to treat
these inhomogeneous DNLSE is introduced by construction of an
explicit transformation which maps the VCDNLSE to the DNLSE, which
has been proven to be effective. What is more, rogue waves have
recently been the subject of intensive investigations such as very
interesting "discussion \& debate" \cite{Pelinovsky1,Ruban} on the
fundamental definition and possible applications, a numerical study
of the evolution dynamics of optical rogue waves\cite{Dudley},
researches of evolution of rogue waves in interacting wave systems
\cite{Gronlund} and in microwave system\cite{Hohmann}, a minimal
model\cite{lai} for the emergence and statistical characterization
of robust complex branched wave patterns(or rogue waves) in optical
media. One of the possible formation mechanisms for rogue waves are
rational solutions, the creation of
 breathers that appear due to modulation
instability\cite{Peregrine,Dysthe,Akhmediev1,Ankiewicz,matveevb,Shrira},
so the nature of rational solutions of these inhomogeneous DNLSE deserves
investigation from consideration of optical rogue waves originating from one
key ingredients:inhomogeneity\cite{Arecchi}. However,the model
equation of a system with inhomogeneity is the VCDNLSE, which is associated
with a non-isospectral problem.Usually, it is intricate to solve non-isospectral problem
 directly. So we show an explicit transformation between the VCDNLSE and the DNLSE,
 which give two kinds of rogue waves of the VCDNLSE from the rogue wave and rational
 traveling solution  of the DNLSE.

  The organization of this paper is as follows. In section 2, a general transformation which maps
the VCDNLSE to the DNLSE is given with several arbitrary functions.
These arbitrary functions provide a possibility to design
interesting integrable model. In section 3, the solutions of the
VCDNLSE,especially rogue waves, are discussed according to analytic
solutions of the DNLSE by means of our general transformation.The
conclusion will be given in section 4.

\section{General Method}
Consider the VCDNLSE of the DNLSE type:
\begin{equation}\label{VCDNLS}
i\psi_{t}+\beta\psi_{xx}+K\psi^2\psi^\ast+i\gamma(\psi^2\psi^\ast)_{x}+i\Gamma\psi=0,
\end{equation}
where $\beta= \beta(x,t),K = K(x, t), \gamma = \gamma(x, t)
,\Gamma=\Gamma(x,t)$ are four real functions of x and t. First focus
on the integrability of eq.(\ref{VCDNLS}) by means of looking for a direct
transformation between the VCDNLSE and the DNLSE,
\begin{equation}\label{dnlsI1}
iQ_{T}+e~{Q_{XX}}+i(Q^2Q^\ast)_{X}=0,
\end{equation}
where e =±1, Q= Q(X, T). Moreover, coefficients $\beta,K , \gamma
 ,\Gamma$ are given analytically by this
transformation. Therefore, one advantage of this transformation is
to solve the VCDNLSE by using all known solutions of the DNLSE as we
discussed in the above section.

  To this purpose, a trial transformation\begin{equation}\label{transformation}
\psi(x,t)=Q(X,T)p(x,t)e^{i\phi(x,t)}
\end{equation}
is introduced with X = X(x, t), T = T(t). So the central task is to
determine the concrete expressions of real smooth functions
{$\beta,K , \gamma
 ,\Gamma,X, p, \phi(x,t)$ } by requesting Q(X, T) to satisfy the standard DNLSE
eq.(\ref{dnlsI1}). By substituting the transformation eq.(\ref{transformation}) into eq.(\ref{VCDNLS}),and
setting \begin{equation}\label{ttiaojian1}
p(x,t)=\dfrac{1}{T_{t}},\beta(x,t)=e\dfrac
{T_{t}}{X_{x}^2},\gamma=\dfrac{T_{t}^3}{X_{x}}
\end{equation}
without loss of generality of the transformation, then it
becomes\begin{equation}\label{transformation1}
iQ_{T}+e~Q_{XX}+i(Q^2Q^\ast)_{X}-Q^2Q^\ast\dfrac{k_1}{{T_t}^3{X_x}}+Q\dfrac{ik_2+k_3}{{T_t}^2{X_x}^2}+Q_{X}\dfrac{ik_4+k_5}{{T_t}{X_x}^2}=0
\end{equation}
Note $Q_X$ denotes $\dfrac{\partial Q}{\partial X}$, $T_t$ denotes
$\dfrac{d T}{d t}$ and so on. Here $k_i$, $i=1, 2, 3,4,5$, are given
by \begin{align} &k_1= {T_{t}}^3\phi_{x}-K~X_{x},\ \  k_2= e~\phi_{xx}{T_{t}}^2+\Gamma{T_{t}{X_{x}}^2}-T_{tt}{X_{x}}^2, \nonumber \\
 &k_3=-\phi_{t}T_{t}X_{x}^2-e~{\phi_{x}}^2{T_{t}}^2,\ \ k_4={X_{x}}^2X_{t}+2e\phi_{x}T_{t}X_{x},\ \ k_5=e~X_{xx}T_{t}.\nonumber
\end{align}
 Obviously,
let $ k_1=k_2=k_3=k_4=k_5=0$, then eq.(\ref{transformation1})
becomes the standard DNLSE. Moreover, we get
\begin{align}
&X=X(x,t)= c_1~x+c_2~x~T+c_3+c_4~T,  \quad T=T(t), \label{tiaojian2}\\
 &\phi=\phi(x,t)=\dfrac{(-c_2^2~T-c_1~c_2)}{4~e}x^2+\dfrac{(-c_4~c_2~T-c_1~c_4)}{2~e}x+\dfrac{-c_4^2~T}{4~e}+c_5 , \label{tiaojian3} \\
&K=-\dfrac{1}{2}\dfrac{{T_{t}^3}(c_2~x+c_4)}{e},\ \ \Gamma=\dfrac{1}{2}\dfrac{c_2~{T_{t}}^2+2~T_{tt}~c_1+2~T_{tt}~T~c_2}{T_{t}(c_1+c_2~T)},\label{tiaojian5}
\end{align}
from $ k_1=k_2=k_3=k_4=k_5=0$ by a  tedious calculation.
Furthermore, taking  $X$ back into eq.(\ref{ttiaojian1}), then
\begin{equation}
p(x,t)=\dfrac{1}{ T_t}, \quad
\beta(x,t)=\dfrac{eT_t}{(c_1+Tc_2)^2},\quad
\gamma=\dfrac{{T_t}^3}{c_1+Tc_2}. \label{tiaojian1}
\end{equation}
So the transformation  in  eq.(\ref{transformation}) indeed maps the
VCDNLSE to the DNLSE as we wanted, which is determined by
eq.(\ref{VCDNLS}) to eq.(\ref{dnlsI1})
through five  arbitrary
 constants  $c_1,c_2, c_3, c_4, c_5$ ($c_1^{2}+c_2^{2}\neq0$) and one real arbitrary
 function $T(t)$.
 Lastly, we would like to point out that  we may set $X=X(x,t)$ and  $T=T(x,t)$ in transformation
 eq.(\ref{transformation}) for better universality. However, to eliminate the term $(\dfrac{\partial^2}{\partial
 X\partial T} Q(X,T))$ in the transformed equation, we have to set $(\dfrac{\partial }{\partial x} X )=0$ or
$(\dfrac{\partial }{\partial x} T )=0$. So we choose directly
$T=T(t)$ for simplicity in eq.(\ref{transformation}). Thus we can
design $\beta,K , \gamma ,\Gamma$ according to different physical
considerations by means of the selections of the arbitrary constants
and functions in the transformations above, such that the
integrability is guaranteed. Therefore, we call that the VCDNLSE
possesses the designable integrability(DI)\cite{he}, which
originates from the rigid integrability
 of the DNLSE and the transformation.

  As the end of this section, we would  show several examples
of nonautonomous DNLSE eq.(\ref{VCDNLS}). Their solutions can be
obtained from known solutions of the DNLSE by the transformation in  eq.(\ref{transformation}).\\
\noindent {\bf Example 1.} Taking  $c_1=1,c_2=0,c_3=0,c_4=0,c_5=0,e=-1$ and
$T=\dfrac{1}{2}\cos(t)+t$ back into  eq.(\ref{tiaojian5}) and eq.(\ref{tiaojian1}),we can get the following
equation \begin{equation}\label{EQ1}
i\psi_{t}-(-\dfrac{1}{2}\sin(t)+1)\psi_{xx}+i\dfrac{1}{8}(\sin(t)-2)^{3}(\psi^2\psi^\ast)_{x}+i\dfrac{\cos(t)}{\sin(t)-2}\psi=0
\end{equation} the following items are given for the equation: (i)the trajectory is not a straight line when it evolves with
time t and the profile is variable.(ii) It is interesting to get the
rogue wave
of the eq.(\ref{EQ1}).\\
\noindent {\bf Example 2.}  Taking  $c_1=3,c_2=0,c_3=0,c_4=0,c_5=0,e=-1$ and
$T=\dfrac{1}{4} t^{3}+\dfrac{3}{4}t$ back into  eq.(\ref{tiaojian5}) and eq.(\ref{tiaojian1}),we
can get the following equation \begin{equation}\label{EQ2}
i\psi_{t}-\dfrac{t^2+1}{12}\psi_{xx}-i\dfrac{9(t^{2}+1)^{3}}{64}(\psi^2\psi^\ast)_{x}+i\dfrac{2t}{t^2+1}\psi=0
\end{equation}The two kinds of rogue waves of the eq.(\ref{EQ2})
will be  given in next section.

\section{two kinds of rogue waves of vcdnlse}
The rational solutions and rogue waves of eq.(\ref{dnlsI})\cite{xuhe1} are
 given in the following.
 \begin{equation} \label{rational1}
Q_{rational}=4 \beta_{1} \exp{(2 i {\beta_{1}}^{2}(-X +2  {\beta_{1}}^{2}T))}
\dfrac{(4i{\beta_{1}}^{2}(4{\beta_{1}}^{2} T-X)-1)^{3}}{(16{\beta_{1}}^{4}({4\beta_{1}}^{2}T-X)^2+1)^2},
\end{equation}
with an arbitrary real constant $\beta_{1}$. Obviously, the rational solution is a line soliton,
and its trajectory is defined explicitly by
\begin{equation}\label{rational2} X=4 {\beta_{1}}^{2} T,
 \end{equation}
on $(x-t)$ plane.
Here
\begin{equation}\label{rationaljie}
Q_{rogue~wave}=\dfrac{r_1 r_2 r_3}{r_4 r_5},
\end{equation}
\begin{eqnarray*}
&&r_1=2 \exp(2i(\alpha_1^2+\beta_1^2)(2 T \alpha_1^2+X-2 T \beta_1^2)),\nonumber\\
&&r_2=\beta_1(16 \beta_1^2 \alpha_1^2 (4 T \alpha_1^2+X)^2+16 \beta_1^4 (4T\beta_1^2-X)^2+8i \beta_1^2 (X+4 T \alpha_1^2-8 T \beta_1^2)+1),\nonumber\\
&&r_3=\lefteqn{2(16 \beta_1^2\alpha_1^2 (4 T \alpha_1^2+X)^2+16\beta_1^4 (4 T\beta_1^2-X)^2-8\alpha_1\beta_1(X+4 T \alpha_1^2-8 T \beta_1^2)+1){}}\nonumber\\
&&{}\mbox{\hspace{0.8cm}}\times(-\alpha_1+16\beta_1(\beta_1^4-\alpha_1^4) T-4\beta_1(\alpha_1^2+\beta_1^2) X+16i \alpha_1\beta_1^2(\alpha_1^2+\beta_1^2) T-i\beta_1)\nonumber\\
&&{}\mbox{\hspace{0.8cm}}-(16\beta_1^2\alpha_1^2 (4 T\alpha_1^2+X)^2+16\beta_1^4(4 T \beta_1^2-X)^2+8i\beta_1^2(X+4 T \alpha_1^2-8 T\beta_1^2)+1)\nonumber\\
&&{}\mbox{\hspace{0.8cm}}\times(\alpha_1+16\beta_1(\beta_1^4-\alpha_1^4)T-4\beta_1(\alpha_1^2+\beta_1^2) X+16i\alpha_1\beta_1^2(\alpha_1^2+\beta_1^2)T+\beta_1i),\nonumber\\
&&r_4=\alpha_1+16\beta_1(\beta_1^4-\alpha_1^4)T-4\beta_1(\alpha_1^2+\beta_1^2)X+16 i\alpha_1\beta_1^2(\alpha_1^2+\beta_1^2)T+\beta_1i,\nonumber\\
&&r_5=(-16\beta_1^2\alpha_1^2(4 T\alpha_1^2+X)^2-16\beta_1^4(4 T \beta_1^2-X)^2+8i\beta_1^2(X+4 T\alpha_1^2-8 t\beta_1^2)-1)^2.\nonumber\\
\end{eqnarray*}
By letting $ X \rightarrow {\infty}, \ T \rightarrow {\infty}$,
 so $|Q_{rogue~wave}|^{2}\rightarrow 4\beta_{1}^2$.
The maximum amplitude of $|Q_{rogue~wave}|^{2}$ occurs at $T = 0$ and $X=0$ and is
 equal to $36\beta_{1}^{2}$, and the minimum amplitude of $|Q_{rogue~wave}|^{2}$ occurs at
  $T = \pm\dfrac{3}{16\sqrt{3(4\alpha_{1}^{2}+\beta_{1}^{2})}\beta_{1}(\alpha_{1}^{2}+\beta_{1}^{2})}$ and
  $X=\mp\dfrac{9\alpha_{1}^{2}}{4\sqrt{3(4\alpha_{1}^{2}+\beta_{1}^{2})}
  \beta_{1}(\alpha_{1}^{2}+\beta_{1}^{2})}$ and is equal to $0$.

The transformation eq.(\ref{transformation}) provides an efficient way to construct
exact and analytic solutions $\psi(x, t)$ of the designable VCDNLSE
from known solutions $Q(X, T)$ of the DNLSE, such that we can
explore the dynamical evolution of solitons of the VCDNLSE
conveniently. For example, setting a rational soliton $Q_{rational}$
which is given by eq.(\ref{rational1}).

Obviously,the profile of the rational soliton $Q_{rational}$ on the plane
of (X,T) is invariant when soliton evolves with time $T$. A rational
soliton of the VCDNLSE,\begin{equation} |\psi(x,t)|=4|p(x,t)||\beta_{1}|
\dfrac{1}{(1+16 \beta_{1}^{4} (-X+4 \beta_{1}^{2} T)^{2})^{1/2}}
\end{equation}
is obtained by using transformation eq.(\ref{transformation}). Note that the profile of
the soliton $|\psi(x,t)|$ of the VCDNLSE is not preserved when it
evolves with time t because the profile $|p(x, t)|$ is a function of
t and the trajectory is a curve $x = x(t)$ on the plane of $(x,t)$,
which is defined implicitly by
\begin{equation}
X=4 {\beta_{1}}^{2} T
\end{equation}
This shows that the profile of the soliton of the VCDNLSE is
designable by using different $\beta= \beta(t),K = K(x,t), \gamma =
\gamma(t) ,\Gamma=\Gamma(t)$, which can be realized by choosing
arbitrary constants  $c_1,c_2, c_3, c_4,$\\$ c_5$ ($c_1^{2}+c_2^{2}\neq0$)
and one real arbitrary function $T(t)$ in transformation eq.(\ref{transformation}). In
particular, as we shall show in the following example the profile of
$\psi(x, t)$ is dependent of t , because the $p = p(t)$
is t-dependent. To further illustrate the property of dynamical evolution of the
soliton $|\psi(x, t)|$ of the VCDNLSE,two example is given here.\\

\noindent {\bf Case 1).} According to the above formula
eq.(\ref{transformation}), the solution of the eq.(\ref{EQ1})  is
given by taking  $c_1=1, c_2=0, c_3=0, c_4=0, c_5=0, e=-1$ and
$T=\dfrac{1}{2}\cos(t)+t $ back into eq.(\ref{tiaojian2}),
eq.(\ref{tiaojian3}) and eq.(\ref{tiaojian1}),which is a deformed
rational solution.  (a) The dynamical evolution of the solution
generated  from eq.(\ref{rational1}) with specific parameters
$\beta_{1}=\dfrac{1}{2}$ is plotted in Figure 1, and we also give
its trajectory of the wave  by Figure 2. (b) The dynamical evolution
of the solution generated by the rogue wave eq.(\ref{rationaljie})
with specific parameters
$\alpha_{1}=\dfrac{1}{2},\beta_{1}=\dfrac{1}{2}$ is plotted in Figure 3.\\

\noindent {\bf Case 2).} According to the above formula
eq.(\ref{transformation}),the solution of the eq.(\ref{EQ2})  is
gived by taking  $c_1=3, c_2=0, c_3=0, c_4=0, c_5=0, e=-1$ and $T=\dfrac{1}{4}
t^{3}+\dfrac{3}{4}t$ back into eq.(\ref{tiaojian2}),
eq.(\ref{tiaojian3}) and eq.(\ref{tiaojian1}), which is a deformed
rational solution as case $1)$.     (a) The dynamical evolution of
the solution generated from the rational traveling solution
eq.(\ref{rational1}) with specific parameters
$\beta_{1}=\dfrac{1}{2}$ is plotted in Figure 4. (b) The dynamical
evolution of the solution generated from the rogue wave
eq.(\ref{rationaljie}) with specific parameters
$\alpha_{1}=\dfrac{1}{2},\beta_{1}=\dfrac{1}{2}$ is plotted in
Figure 5.

Figure 1 and Figure 2 shows intuitively two interesting features of
the solutions of the VCDNLSE: (i) The trajectory is not a straight
line when it evolves with time t, as we pointed in the above
paragraph. (ii) The profile is variable.This is also supported
visibly by Fig.3, Fig.4 and Fig.5. What is more,we get two kinds of
rogue waves of the VCDNLSE by an explicit transformation,they are
plotted in Fig.4 and Fig.5. The first kind of rogue wave has
vanishing boundary condition, and the second non-vanishing boundary
condition.With the inhomogeneous condition, there are remarkable
differences between the VCDNLSE and the DNLSE. Note that there are
many other kinds of solutions of the DNLSE, such as dark
soliton, positon, periodic solution,etc., which can be used to
generate the solutions of the VCDNLSE. In our examples, merely the
rational solution of the DNLSE is used. Especially, from the
“seed” of the multi-soliton solutions of the DNLSE, the multi-soliton
solutions of the VCDNLSE can be obtained, and then its interaction
properties are discussed. However, it is very difficult to find
multi-soliton solutions of the VCDNLSE from the view of
non-isospectral integrable system.

Very recently, the experiment results\cite{Arecchi} show that the
inhomogeneity of the medium is one kind of crucial ingredients to
generate optical rogue wave. Note that the inhomogeneity is usually
represented by the variable coefficients in the integrable
equations.  The rogue wave in case 2 a) is given from a rational
traveling wave eq.(\ref{rational1}) which shows the direct
contribution of the inhomogeneity  to generate it. Moreover, we can
design different inhomogeneity by selecting different arbitrary
functions in transformation, eq.(\ref{transformation}), to get rogue
wave from known solutions of the DNLSE. At last,in order to get more
theoretical support of the same experiment work\cite{Arecchi}, we
shall further explore which kinds of the inhomogeneity (or variable
coefficients) can generate rogue wave.

\section{Conclusions}
One kind of explicit transformation between DNLSE and VCDNLSE is
introduced, which shows that VCDNLSE eq.(\ref{VCDNLS}) with special
coefficients $\beta,K, \gamma,\Gamma$ from different physical
considerations has DI. The profile of its solutions including the
rational solutions and others are also tunable intentionally by
using different $p(x, t), X(x, t)$ and  $T(t),$ which can be
realized by choosing different arbitrary functions. The results in
this paper show that some unusual behaviors of solution for the
VCDNLSE originate from the usual DNLSE .There are two kinds of rogue
waves of the VCDNLSE by an explicit transformation.One kind of rogue
wave has vanishing boundary condition, and the other non-vanishing
boundary condition. Those results are useful to study the properties
of the inhomogeneous space, and show that the DI is an important
feature of the VCDNLSE. Further, the DI of the VCDNLSE provides a
way to get different profiles of the rogue waves  by choosing
arbitrary functions in transformation, and thus also provides a
method to control the rogue waves in inhomogeneous systems in
physical experiment and theory analysis.

  {\bf Acknowledgments} {\noindent \small  This work is supported by
the NSF of China under Grant No.10971109 and K.C.Wong Magna Fund in
Ningbo University. Jingsong He is also supported by Program for NCET
under Grant No.NCET-08-0515. Shuwei Xu is also supported by the
Scientific Research Foundation of Graduate School of Ningbo
University. We thank Prof. Yishen Li (USTC,Hefei, China) for his
useful suggestions on the rogue wave. We also thank
Prof. Robert von Fay-Siebenburgen (University of Sheffield)
for his valuable discussions on the possible application of the rogue wave of this paper
during his visit to Ningbo.}

\begin{figure}[htbp]     \centering \mbox{}\hspace{-1cm}
      \includegraphics[width=.3\textwidth]{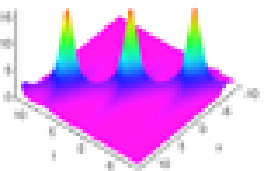}
      \mbox{}\vspace{1cm}\caption{
The dynamical evolution of the solution is generated by the rational
traveling solution with specific parameters $\beta_{1}=\dfrac{1}{2}$
from case 1 a).}
\end{figure}

\begin{figure}[htbp]     \centering \mbox{}\hspace{1cm}
      \includegraphics[width=.3\textwidth]{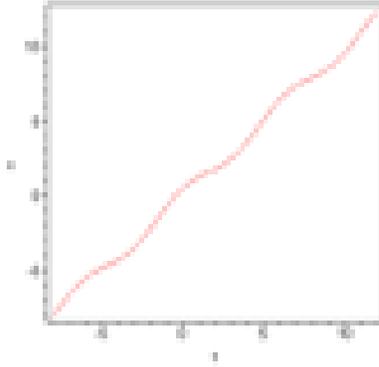}
      \mbox{}\vspace{-0.5cm}\caption{
The dynamical evolution trajectory is defined explicitly by $x-\dfrac{1}{2}cos(t)-t=0$
from case 1 a).}
\end{figure}

\begin{figure}[htbp]     \centering \mbox{}\hspace{1cm}
      \includegraphics[width=.3\textwidth]{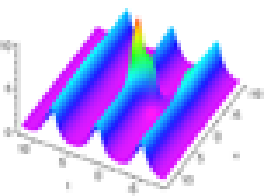}
      \mbox{}\vspace{-0.35cm}\caption{
The dynamical evolution of the solution is generated by the rogue
wave with specific parameters
$\alpha_{1}=\dfrac{1}{2},\beta_{1}=\dfrac{1}{2}$ from case 1 b). }
\end{figure}

\begin{figure}[htbp]     \centering \mbox{}\hspace{1cm}
      \includegraphics[width=.3\textwidth]{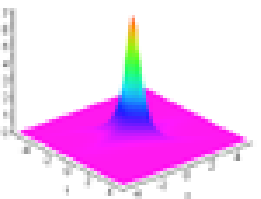}
      \mbox{}\vspace{-0.35cm}\caption{
The dynamical evolution of the solution is generated by the rational
traveling solution with specific parameters $\beta_{1}=\dfrac{1}{2}$
from case 2 a).  }
\end{figure}

\begin{figure}[htbp]     \centering \mbox{}\hspace{1cm}
      \includegraphics[width=.3\textwidth]{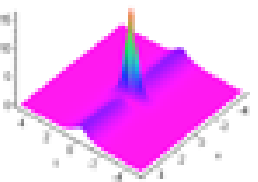}
      \mbox{}\vspace{-0.35cm}\caption{
The dynamical evolution of the solution is generated by the rogue
wave with specific parameters
$\alpha_{1}=\dfrac{1}{2},\beta_{1}=\dfrac{1}{2}$ from case 2 b).  }
\end{figure}

\end{document}